\documentclass[twocolumn,showpacs,preprintnumbers,pra,hyperref]{revtex4}
\usepackage{amssymb}
\usepackage{amsfonts}
\usepackage{amsmath}
\usepackage{graphicx}

\begin{document}

\title{An optical diode and magnifier from a general function photonic crystals}
\author{Xiang-Yao Wu$^{a}$\thanks{%
E-mail: wuxy2066@163.com}, Bo-Jun Zhang$^{a}$, Xiao-Jing Liu$^{a}$, Si-Qi Zhang$^{a}$\\
 Jing Wang$^{a}$, Nuo Ba$^{a}$, Zhong Hua$^{a}$ and Jing-Wu Li$^{b}$} \affiliation{$^{a}${\small
Institute of Physics, Jilin Normal University, Siping 136000, China}\\
$^{b}${\small  Institute of Physics, Xuzhou Normal University,
Xuzhou 221000, China} }

\begin{abstract}
We have presented a new general function photonic crystals
(GFPCs), which refractive index is a function of space position.
Based on Fermat principle, we achieve the motion equations of
light in one-dimensional general function photonic crystals, and
calculate its transfer matrix. In this paper, we choose the line
refractive index function for two mediums $A$ and $B$, and obtain
new results: (1) when the line function of refractive indexes is
up or down, the transmissivity can be far larger or smaller than
$1$. (2) when the refractive indexes function increase or decrease
at the direction of incident light, the light intensity should be
magnified or weaken, which can be made light magnifier or
attenuator. (3) The GFPCs can also be made optical diode. The new
general function photonic crystals can
be applied to design more optical instruments.\\

\end{abstract}

\maketitle

\maketitle {\bf 1. Introduction} \vskip 8pt

Photonic crystals are artificial materials with periodic
variations in refractive index that are designed to affect the
propagation of light [1-4]. An important feature of the photonic
crystals is that there are allowed and forbidden ranges of
frequencies at which light propagates in the direction of index
periodicity. Due to the forbidden frequency range, known as
photonic band gap (PBG) [5-6], which forbids the radiation
propagation in a specific range of frequencies. The existence of
PBGs will lead to many interesting phenomena, e.g., modification
of spontaneous emission [7-9] and photon localization [10]. Thus
numerous applications of photonic crystals have been proposed in
improving the performance of optoelectronic and microwave devices
such as high-efficiency semiconductor lasers, right emitting
diodes, wave guides, optical filters, high-Q resonators, antennas,
frequency-selective surface, optical limiters and amplifiers
[11-18].

In Ref. [19], we have proposed special function photonic crystals,
which the medium refractive index is the function of space
position, but the function value of refractive index is equal at
two endpoints of every medium $A$ and $B$. In this paper, we
present a new general function photonic crystals (GFPCs), which
refractive index is a arbitrary function of space position. Unlike
conventional photonic crystals (PCs), which structure grow from
two materials, A and B, with different dielectric constants
$\varepsilon_{A}$ and $\varepsilon_{B}$. Firstly, we give the
motion equation of light in one-dimensional GFPCs based on Fermat
principle. Secondly, we calculate the transfer matrix for the
GFPCs, which is different from the transfer matrix of the
conventional PCs. Thirdly, we give the band gap structure and
transmissivity. Finally, we choose the linearity refractive index
functions for two  medium $A$ and $B$, and give the light field
distribution in the GFPCs. We obtain some new results: (1) when
the line function of refractive indexes is up, the transmissivity
can be far larger than $1$. (2) when the line function of
refractive indexes is down, the transmissivity can be far smaller
than $1$. (3) when the refractive indexes function increase at the
incident direction of light, the light intensity should be
magnified, which can be made light magnifier. (4) when the
refractive indexes function decrease at the incident direction of
light, the light intensity should be weaken, which can be made
light attenuator. (5) The GFPCs can be made optical diode, which
transmits light from an input to an output, but not in reverse
direction.

\vskip 8pt

{\bf 2. The light motion equation in general function photonic
crystals} \vskip 8pt

For the general function photonic crystals, the medium refractive
index is a periodic function of the space position, which can be
written as $n(z)$, $n(x, z)$ and $n(x, y, z)$ corresponding to
one-dimensional, two-dimensional and three-dimensional function
photonic crystals. In the following, we shall deduce the light
motion equations of the one-dimensional general function photonic
crystals, i.e., the refractive index function is $n=n(z)$,
meanwhile motion path is on $xz$ plane. The incident light wave
strikes plane interface point $A$, the curves $AB$ and $BC$ are
the path of incident and reflected light respectively, and they
are shown in FIG. 1.
\begin{figure}[tbp]
\includegraphics[width=8.5 cm]{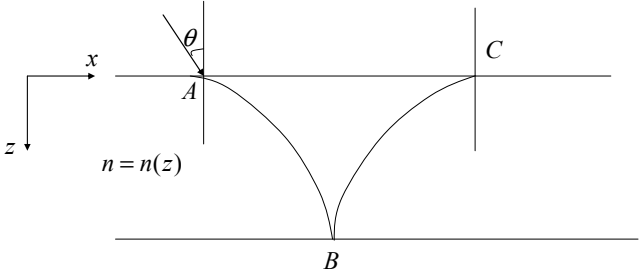}
\caption{The motion path of light in the medium of refractive
index $n(z)$.}
\end{figure}

The light motion equation can be obtained by Fermat principle, it
is
\begin{eqnarray}
\delta\int^{B}_{A}n(z) ds=0.\label{1}
\end{eqnarray}
In the two-dimensional transmission space, the line element $ds$
is
\begin{eqnarray}
ds=\sqrt{(dx)^{2}+(dz)^{2}}=\sqrt{1+\dot{z}^{2}}dx,
\end{eqnarray}
where $\dot{z}=\frac{dz}{dx}$, then Eq. (1) becomes
\begin{eqnarray}
\delta\int^{B}_{A}n(z)\sqrt{1+(\dot{z})^{2}}dx=0.
\end{eqnarray}
The Eq. (3) change into
\begin{eqnarray}
\int^{B}_{A}(\frac{\partial(n(z)\sqrt{1+\dot{z}^{2}})}{\partial
z}\delta z+\frac{\partial(n(z)\sqrt{1+\dot{z}^{2}})}{\partial
\dot{z}}\delta\dot{z})dx=0,
\end{eqnarray}
At the two end points $A$ and $B$, their variation is zero, i.e.,
$\delta z (A)=\delta z (B)=0$. For arbitrary variation $\delta z$, the Eq. (4) becomes \\
\begin{eqnarray}
&&\frac{dn(z)}{dz}\sqrt{1+\dot{z}^{2}} -\frac{d n(z)}{d
z}\dot{z}^{2}(1+\dot{z}^{2})^{-\frac{1}{2}}
\nonumber\\&&-n(z)\frac{\ddot{z}\sqrt{1+\dot{z}^{2}}
-\dot{z}^{2}\ddot{z}(1+\dot{z}^{2})^{-\frac{1}{2}}}{1+\dot{z}^{2}}
=0,
\end{eqnarray}
simplify Eq. (5), we have
\begin{eqnarray}
\frac{d n(z)}{n(z)} = \frac{\dot{z}d\dot{z}}{1+\dot{z}^{2}}.
\end{eqnarray}\\
The Eq. (6) is light motion equation in one-dimensional function
photonic crystals.
\begin{figure}[tbp]
\includegraphics[width=8.5 cm]{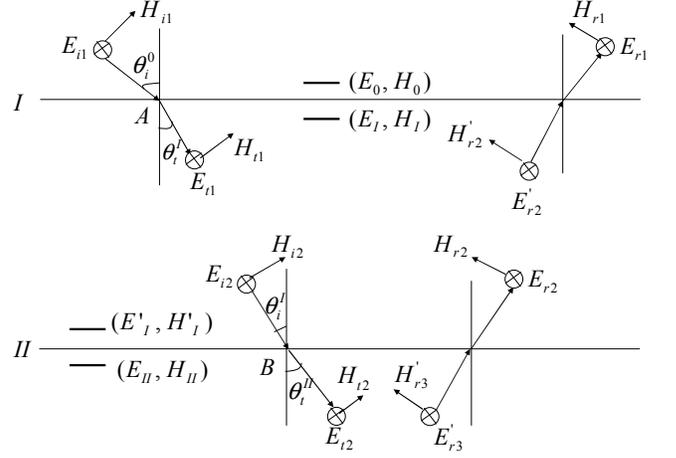}
\caption{The light transmission and electric magnetic field
distribution figure in FIG.1 medium.} \label{Fig1}
\end{figure}
\vskip 8pt

{\bf 3. The transfer matrix of one-dimensional general function
photonic crystals} \vskip 8pt

In this section, we should calculate the transfer matrix of
one-dimensional general function photonic crystals. In fact, there
is the reflection and refraction of light at a plane surface of
two media with different dielectric properties. The dynamic
properties of the electric field and magnetic field are contained
in the boundary conditions: normal components of $D$ and $B$ are
continuous; tangential components of $E$ and $H$ are continuous.
We consider the electric field perpendicular to the plane of
incidence, and the coordinate system and symbols as shown in FIG.
2.

On the two sides of interface I, the tangential components of
electric field $E$ and magnetic field $H$ are continuous, there
are

\begin{eqnarray}
\left \{ \begin{array}{ll}
 E_{0}=E_{I}=E_{t1}+E'_{r2}\\
H_{0}=H_{I}=H_{t1}\cos\theta_{t}^{I}-H'_{r2}\cos\theta_{t}^{I}.
\end{array}
\right.
\end{eqnarray}
On the two sides of interface II, the tangential components of
electric field $E$ and magnetic field $H$ are continuous, and give
\begin{eqnarray}
\left \{ \begin{array}{ll}
 E_{II}=E'_{I}=E_{i2}+E_{r2}\\
H_{II}=H'_{I}=H_{i2}\cos\theta_{i}^{I}-H_{r2}\cos\theta_{i}^{I},
\end{array}
\right.
\end{eqnarray}
the electric field ${E_{t1}}$ is
\begin{eqnarray}
E_{t1}=E_{t10}{e^{i(k_{x}x_{A}+k_{z}z)}|_{z=0}}=E_{t10}e^{i\frac{\omega}{c}n(0)\sin\theta_{t}^{I}x_{A}},
\end{eqnarray}
and the electric field ${E_{i2}}$ is
\begin{eqnarray}
E_{i2}&=&E_{t10}{e^{i(k'_{x}x_{B}+k'_{z} z)}|_{z=b}}
\nonumber\\&=&E_{t10}e^{i\frac{\omega}{c}n(b)(\sin\theta_{i}^{I}x_{B}+\cos\theta_{i}^{I}
b)}.
\end{eqnarray}
Where $x_{A}$ and $x_{B}$ are $x$ component coordinates
corresponding to point $A$ and point $B$. We should give the
relation between $E_{i2}$ and $E_{t1}$. By integrating the two
sides of Eq. (6), we can obtain the coordinate component $x_{B}$
of point $B$
\begin{eqnarray}
\int^{n(z)}_{n(0)}\frac{dn(z)}{n(z)}=\int^{k_{z}}_{k_{0}}\frac{\dot{z}d\dot{z}}{1+\dot{z}^{2}},
\end{eqnarray}
to get
\begin{eqnarray}
k_z^2=(1+k_0^2)(\frac{n(z)}{n(0)})^2-1,
\end{eqnarray}
and
\begin{eqnarray}
dx=\frac{dz}{\sqrt{(1+k_{0}^{2})(\frac{n(z)}{n(0)})^{2}-1}}.
\end{eqnarray}
where $k_{0}=\cot\theta_{t}^{I}$ and $k_{z}=\frac{d z}{d x}$ From
Eq. (12), there is $n(z)>n(0)\sin\theta^{I}_{t}$. and the
coordinate $x_{B}$ is
\begin{eqnarray}
x_{B}=x_{A}+\int^{b}_{0}\frac{dz}{\sqrt{(1+k_{0}^{2})(\frac{n(z)}{n(0)})^{2}-1}},
\end{eqnarray}
where $b$ is the medium thickness of FIG. 1 and FIG. 2.\\
By substituting Eqs. (9) and (14)into (10), and using the equality
\begin{eqnarray}
 n(0)\sin\theta_{t}^{I}=n(b)\sin\theta_{i}^{I},
\end{eqnarray}
  we have
\begin{eqnarray}
E_{i2}&=&E_{t1}e^{i{\delta}_{b}},
\end{eqnarray}
where
\begin{figure}[tbp]
\includegraphics[width=8 cm]{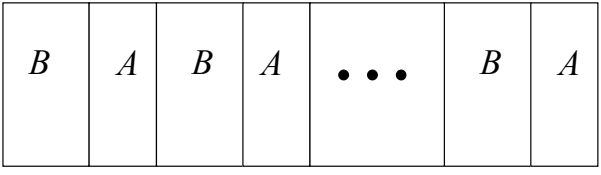}
\caption{The structure $(BA)^{N}$ of the general function photonic
crystals.}
\end{figure}
\begin{eqnarray}
\delta_{b}=\frac{\omega}{c}n_{b}(b)(\cos\theta_{i}^{I}b+\sin\theta_{i}^{I}
\int^{b}_{0}\frac{dz}{\sqrt{\frac{n_{b}^{2}(z)}{n_{0}^{2}\sin^{2}\theta_{i}^{0}}-1}}),
\end{eqnarray}
and similarly
\begin{eqnarray}
E'_{r2}=E_{r2}e^{i\delta_{b}}.
\end{eqnarray}
Substituting Eqs. (16) and (18) into (7) and (8), and using
$H=\sqrt{\frac{\varepsilon_{0}}{\mu_{0}}}nE$, we obtain
\begin{eqnarray}
\left(%
\begin{array}{c}
  E_{I} \\
  H_{I} \\
\end{array}%
\right)&=&M_{B}\left(%
\begin{array}{c}
  E_{II} \\
  H_{II} \\
\end{array}%
\right),
\end{eqnarray}
where
\begin{eqnarray}
M_{B}=\left(%
\begin{array}{cc}
 \cos\delta_{b} & -\frac{i\sin\delta_{b}}{\sqrt{\frac{\varepsilon_{0}}{\mu_{0}}}n_{b}(b)\cos\theta_{i}^{I}} \\
 -in_{b}(0)\sqrt{\frac{\varepsilon_{0}}{\mu_{o}}}\cos\theta_{t}^{I}\sin\delta_{b}
 & \frac{n_{b}(0)\cos\theta_{t}^{I}\cos\delta_{b}}{n_{b}(b)\cos\theta_{i}^{I}}\\
\end{array}%
\right),
\end{eqnarray}
The Eq. (20) is the transfer matrix $M$ in the medium of FIG. 1
and FIG. 2. By refraction law, we can obtain
\begin{eqnarray}
\sin\theta^{I}_{t}=\frac{n_{0}}{n(0)}\sin\theta^{0}_{i},\cos\theta^{I}_{t}
=\sqrt{1-\frac{n_{0}^{2}}{n^{2}(0)}\sin^{2}\theta^{I}_{t}},
\end{eqnarray}
where $n_0$ is air refractive index, and $n(0)=n(z)|_{z=0}$. Using
Eqs. (15) and (21), we can calculate $\cos\theta_{i}^{I}$.

\begin{figure}[tbp]
\includegraphics[width=8 cm]{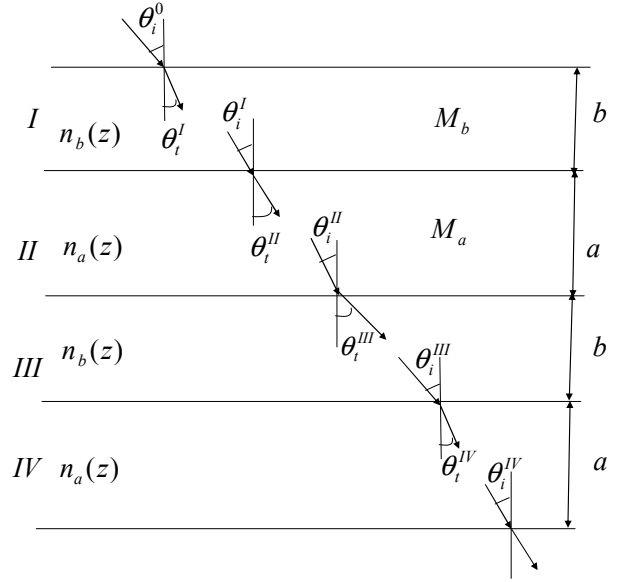}
\caption{The two periods transmission figure of light in general
function photonic crystals.}
\end{figure}
\vskip 8pt

{\bf 4. The transmissivity and light field distribution of
one-dimensional general function photonic crystals} \vskip 8pt

In section 3, we obtain the $M$ matrix of the half period. We know
that the conventional photonic crystals is constituted by two
different refractive index medium, and the refractive indexes are
not continuous on the interface of the two mediums. We could
devise the one-dimensional general function photonic crystals
structure as follows: in the first half period, the refractive
index distributing function of medium $B$ is $n_{b}(z)$. and in
the second half period, the refractive index distributing function
of medium $A$ is $n_{a}(z)$, corresponding thicknesses are $b$ and
$a$, respectively. Their refractive indexes satisfy condition
$n_{b}(b)\neq n_{a}(0)$, their structure are shown in FIG. 3, and
FIG. 4. The Eq. (20) is the half period transfer matrix of medium
$B$. Obviously, the half period transfer matrix of medium A is
\begin{eqnarray}
M_{A}=\left(%
\begin{array}{cc}
 \cos\delta_{a} & -\frac{i\sin\delta_{a}}{\sqrt{\frac{\varepsilon_{0}}{\mu_{0}}}n_{a}(a)\cos\theta_{i}^{II}} \\
 -in_{a}(0)\sqrt{\frac{\varepsilon_{0}}{\mu_{o}}}\cos\theta_{t}^{II}\sin\delta_{a}
 & \frac{n_{a}(0)\cos\theta_{t}^{II}\cos\delta_{a}}{n_{a}(a)\cos\theta_{i}^{II}}\\
\end{array}%
\right),
\end{eqnarray}
where
\begin{eqnarray}
\delta_{a}&=&\frac{\omega}{c}n_{a}(a)[\cos\theta^{II}_{i}\cdot a
\nonumber\\&&
+\sin\theta^{II}_{i}\int^{a}_{0}\frac{dz}{\sqrt{\frac{n_{a}^{2}(z)}{n_{0}^{2}\sin^{2}\theta_{i}^{0}}-1}}],
\end{eqnarray}
\begin{eqnarray}
\cos\theta^{II}_{t}
=\sqrt{1-\frac{n_{0}^{2}}{n_{a}^{2}(0)}\sin^{2}\theta_{i}^{0}},
\end{eqnarray}
and
\begin{eqnarray}
\sin\theta^{II}_{i}=\frac{n_{0}}{n_{a}(a)}\sin\theta_{i}^{0},
\end{eqnarray}
\begin{eqnarray}
\cos\theta^{II}_{i}
=\sqrt{1-\frac{n_{0}^{2}}{n_{a}^{2}(a)}\sin^{2}\theta_{i}^{0}}.
\end{eqnarray}
In one period, the transfer matrix $M$ is
\begin{eqnarray}
&&M=M_{B}\cdot M_{A}\nonumber\\
&&=\left(%
\begin{array}{cc}
  \cos\delta_{b} & \frac{-i\sin\delta_{b}}{\sqrt{\frac{\varepsilon_{0}}{\mu_{0}}}n_{b}(b)\cos\theta_{i}^{I}} \\
 -in_{b}(0)\sqrt{\frac{\varepsilon_{0}}{\mu_{o}}}\cos\theta_{t}^{I}\sin\delta_{b}
 & \frac{n_{b}(0)\cos\theta_{t}^{I}\cos\delta_{b}}{n_{b}(b)\cos\theta_{i}^{I}}\\
\end{array}%
\right) \nonumber\\&&
\left(%
\begin{array}{cc}
   \cos\delta_{a} & \frac{-i\sin\delta_{a}}{\sqrt{\frac{\varepsilon_{0}}{\mu_{0}}}n_{a}(a)\cos\theta_{i}^{II}} \\
 -in_{a}(0)\sqrt{\frac{\varepsilon_{0}}{\mu_{o}}}\cos\theta_{t}^{II}\sin\delta_{a}
 & \frac{n_{a}(0)\cos\theta_{t}^{II}\cos\delta_{a}}{n_{a}(a)\cos\theta_{i}^{II}}\\
\end{array}%
\right).
\end{eqnarray}
The form of the GFPCs transfer matrix $M$ is more complex than the
conventional PCs. The angle $\theta_{t}^{I}$, $\theta_{i}^{I}$,
$\theta_{t}^{II}$ and $\theta_{i}^{II}$ are shown in Fig. 4. The
characteristic equation of GFPCs is
\begin{eqnarray}
\left(%
\begin{array}{c}
  E_{1} \\
  H_{1} \\
\end{array}%
\right)&=&M_{1}M_{2}\cdot\cdot\cdot M_{N}
\left(%
\begin{array}{c}
  E_{N+1} \\
  H_{N+1} \\
\end{array}%
\right) \nonumber\\&=&M_{b}M_{a}M_{b}M_{a}\cdot\cdot\cdot M_{b}M_{a}\left(%
\begin{array}{c}
  E_{N+1} \\
  H_{N+1} \\
\end{array}%
\right)
\nonumber\\&=&M\left(%
\begin{array}{c}
  E_{N+1} \\
  H_{N+1} \\
\end{array}%
\right)=\left(%
\begin{array}{c c}
  A &  B \\
 C &  D \\
\end{array}%
\right)
 \left(%
\begin{array}{c}
  E_{N+1} \\
  H_{N+1} \\
\end{array}%
\right).
\end{eqnarray}
Where $N$ is the period number.
\begin{figure}[tbp]
\includegraphics[width=9 cm]{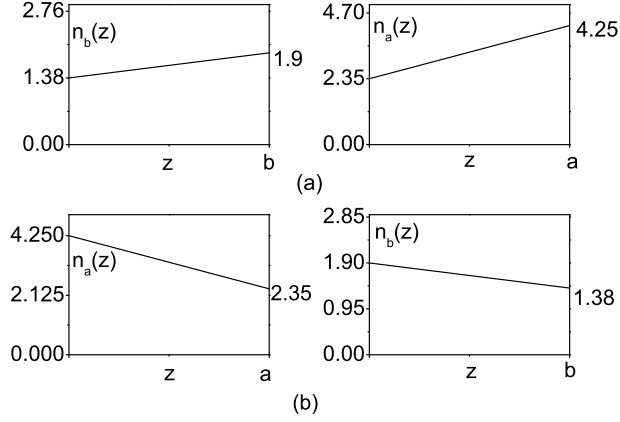}
\caption{The line refractive index functions in a period. The FIG
.5(a) is the up line function of refractive indexes, and FIG .5(b)
is the down line function of refractive indexes.}
\end{figure}
\begin{figure}[tbp]
\includegraphics[width=9 cm]{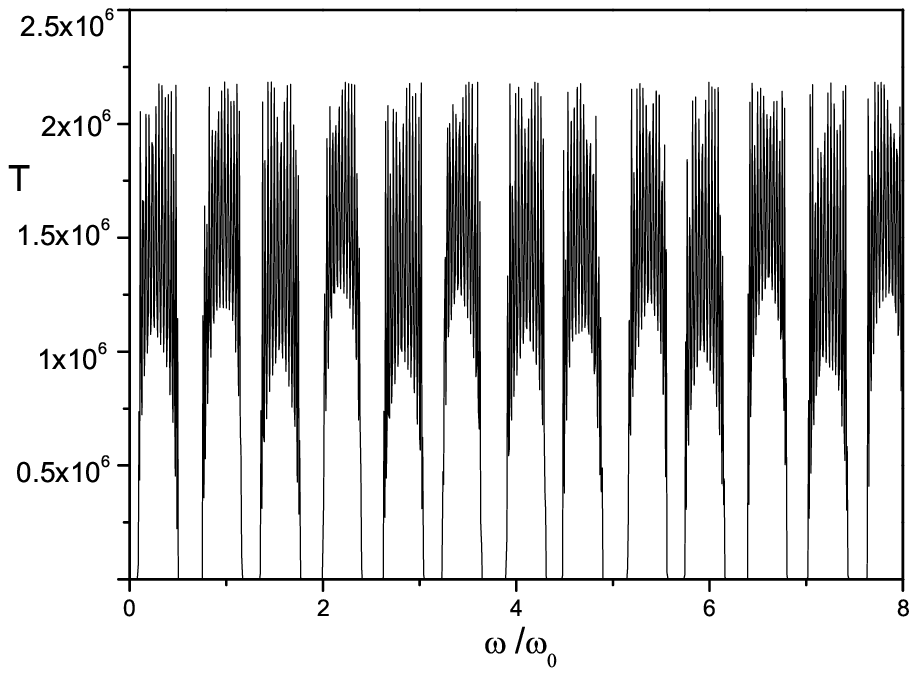}
\caption{The relation between transmissivity and frequency
corresponding to the up line function of refractive indexes (FIG
.5(a))}
\end{figure}
With the transfer matrix $M$ (Eq. (28)), we can obtain the
transmission and reflection coefficient $t$ and $t$, and the
transmissivity and reflectivity $T$ and $R$, they are
\begin{eqnarray}
t=\frac{E_{tN+1}}{E_{i1}}=\frac{2\eta_{0}}{A\eta_{0}+B\eta_{0}\eta_{N+1}+C+D\eta_{N+1}},
\end{eqnarray}
\begin{eqnarray}
r=\frac{E_{r1}}{E_{i1}}=\frac{A\eta_{0}+B\eta_{0}\eta_{N+1}-C-D\eta_{0}}{A\eta_{0}+B\eta_{0}\eta_{N+1}+C+D\eta_{0}},
\end{eqnarray}
and
\begin{eqnarray}
T=t\cdot t^{*},
\end{eqnarray}
\begin{eqnarray}
R=r\cdot r^{*}.
\end{eqnarray}
Where $\eta_{0}=\eta_{N+1}=\sqrt{\frac{\varepsilon_0}{\mu_0}}$. In
the following, we give the electric field distribution of light in
the one-dimensional GFPCs. The propagation figure of light in
one-dimensional GFPCs is shown in FIG. 9. From Eq. (28), we have
\begin{eqnarray}
&&\left(%
\begin{array}{c}
  E_{1} \\
  H_{1} \\
\end{array}%
\right)=M_{1}(d_{1})M_{2}(d_{2})\cdot\cdot\cdot M_{N-1}(d_{N-1})
\nonumber\\&& M_{N}(\Delta z_{N})
\left(%
\begin{array}{c}
  E_{N}(d_{1}+d_{2}{\cdots}+d_{N-1}+\Delta z_{N}) \\
  H_{N}(d_{1}+d_{2}{\cdots}+d_{N-1}+\Delta z_{N}) \\
\end{array}%
\right)
\end{eqnarray}
where $d_1$ and $d_2$ are the thickness of first and second
period, respectively, $\Delta z_{N}$ is the propagation distance
of light in the N-th period, $E_1$ and $H_1$ are the intensity of
incident electric field and magnetic field, and
$E_{N}(d_{1}+d_{2}{\cdots}+d_{N-1}+\Delta z_{N})$ and
$H_{N}(d_{1}+d_{2}{\cdots}+d_{N-1}+\Delta z_{N})$ are the
intensity of the N-th period electric field and magnetic field.
The Eq. (34) can be written as
\begin{eqnarray}
&&\left(%
\begin{array}{c}
  E_{N}(d_{1}+d_{2}{\cdots}+d_{N-1}+\Delta z_{N}) \\
  H_{N}(d_{1}+d_{2}{\cdots}+d_{N-1}+\Delta z_{N}) \\
\end{array}%
\right) =M^{-1}_{N}(\Delta z_{N})\nonumber\\&&
M^{-1}_{N-1}(d_{N-1}) \cdot\cdot\cdot M^{-1}_{2}(d_{2})
M^{-1}_{1}(d_{1})
\left(%
\begin{array}{c}
  E_{1} \\
  H_{1} \\
\end{array}%
\right)\nonumber\\&&\left(%
\begin{array}{cc}
  A(\Delta z_{N}) & B(\Delta z_{N}) \\
  C(\Delta z_{N}) & D(\Delta z_{N}) \\
\end{array}%
\right)\left(%
\begin{array}{c}
  E_{1} \\
  H_{1} \\
\end{array}%
\right),
\end{eqnarray}
the electric field $E_1$ and magnetic field $H_1$ can be written
as
\begin{eqnarray}
E_{1}=E_{i1}+E_{r1}=(1+r)E_{i1},
\end{eqnarray}
\begin{eqnarray}
H_{1}&=&H_{i1}\cos\theta_{i}^{0}-H_{r1}\cos\theta_{i}^{0}
\nonumber\\&=&\sqrt{\frac{\varepsilon_{0}}{\mu_{0}}}\cos\theta_{i}^{0}(1-r)E_{i1}.
\end{eqnarray}
From Eqs. (34)-(36), we can obtain the ratio of the electric field
$E_{N}(d_{1}+d_{2}{\cdots}+d_{N-1}+\Delta z_{N})$ within the GFPCs
to the incident electric field $E_{i1}$, it is
\begin{eqnarray}
&&|\frac{ E_{N}(d_{1}+d_{2}{\cdots}+d_{N-1}+\Delta
z_{N})}{E_{i1}}|^{2}\nonumber\\&&=|A(\Delta z_{N})(1+r)+B(\Delta
z_{N})\sqrt{\frac{\epsilon_{0}}{\mu_{0}}}\cos\theta_{i}^{0}(1-r)|^{2}.
\end{eqnarray}
\vskip 8pt

{\bf 5. Numerical result}

\vskip 8pt

In this section, we report  our numerical results of
transmissivity. We consider refractive indexes of the linearity
functions in a period, it is

\begin{eqnarray}
n_{b}(z)=n_{b}(0)+\frac{n_{b}(b)-n_{b}(0)}{b}z, \hspace{0.1in} 0
\leq z\leq b,
\end{eqnarray}
\begin{eqnarray}
n_{a}(z)=n_{a}(0)+\frac{n_{a}(a)-n_{a}(0)}{a}z, \hspace{0.1in} 0
\leq z\leq a,
\end{eqnarray}
Eqs. (38) and (39) are the line refractive indexes distribution
functions of two half period mediums $B$ and $A$. When the
endpoint values $n_{b}(0)$, $n_{b}(b)$, $n_{a}(0)$ and $n_{a}(a)$
are all given, the line refractive index functions $n_{b}(z)$ and
$n_{a}(z)$ are ascertained. The main parameters are: the half
period thickness $b$ and $a$, the starting point refractive
indexes $n_{b}(0)$ and $n_{a}(0)$, and end point refractive
indexes $n_{b}(b)$ and $n_{a}(a)$, the optical thickness of the
two mediums are equal, i.e., $n_{b}(0)b=n_{a}(0)a$, the incident
angle $\theta_{i}^{0}=0$, the center frequency
$\omega_{0}=1.215\times10^{15} Hz$, the thickness $b=280 nm$,
$a=165 nm$ and the period number $N=16$.

\begin{figure}[tbp]
\includegraphics[width=9 cm]{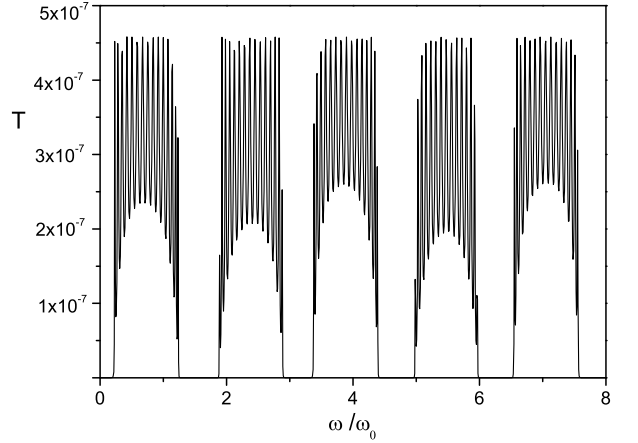}
\caption{The relation between transmissivity and frequency
corresponding to the down line function of refractive indexes (FIG
.5(b)).}
\end{figure}
\begin{figure}[tbp]
\includegraphics[width=9 cm]{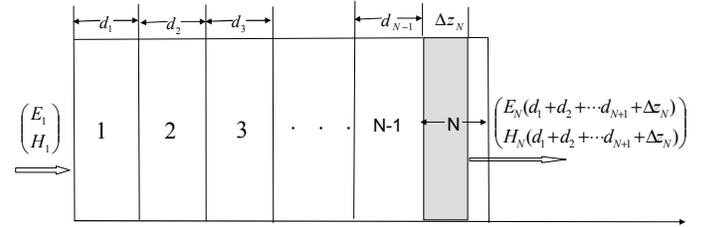}
\caption{The input and output light in the GFPCs.}
\end{figure}
\begin{figure}[tbp]
\includegraphics[width=8.5cm]{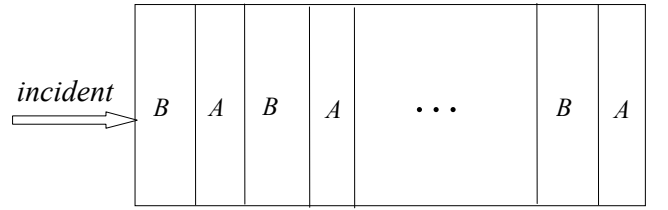}
\caption{The light positive incident to the GFPCs.}
\end{figure}

\begin{figure}[tbp]
\includegraphics[width=8.5 cm]{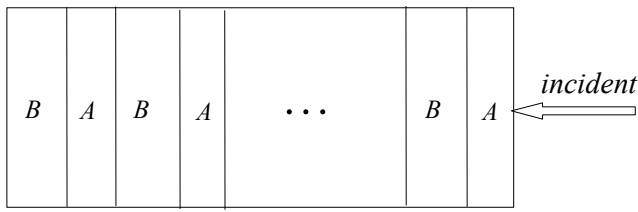}
\caption{The light negative incident to the GFPCs.}
\end{figure}

\begin{figure}[tbp]
\includegraphics[width=8.5 cm]{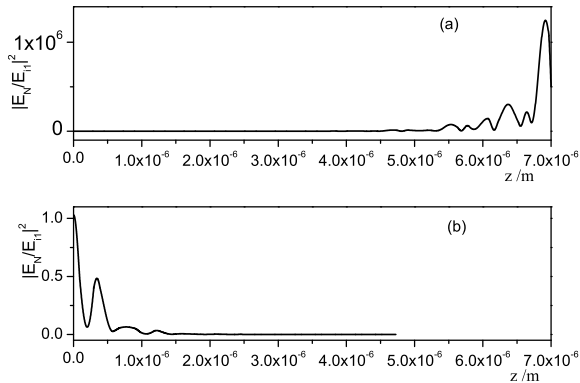}
\caption{The light distribution in the GFPCs. Figure (a) is
corresponding to the positive incident (FIG. 9), and Figure (b) is
corresponding to the positive incident (FIG. 10).}
\end{figure}

In FIG. 5(a), we take $n_{b}(0)=1.38$, $n_{b}(b)=1.9$ for the
medium $B$, and $n_{a}(0)=2.35$, $n_{a}(a)=4.25$ for the medium
$A$, which are the up line function of refractive indexes. In FIG.
5(b), we take $n_{b}(0)=4.25$, $n_{b}(b)=2.35$ for the medium $B$,
and $n_{a}(0)=1.9$, $n_{a}(a)=1.38$ for the medium $A$, which are
the down line function of refractive indexes. By the refractive
indexes function, we can calculate the transmissivity. With the
FIG. 5(a) and 5(b), we obtain the transmissivity distribution in
FIG. (6) and FIG. (7). From FIG. (6) and FIG. (7), we obtain the
results: (1) when the line function of refractive indexes is up,
the transmissivity can be far larger than $1$ ($T$ maximum is
$10^{6}$). (2) when the line function of refractive indexes is
down, the transmissivity can be far smaller than $1$ ($T$ maximum
is $10^{-7}$). In the following, we shall study the light field
distribution of the one-dimensional GFPCs for the light of
positive and negative incident. The positive incident is shown in
FIG. 9 and the negative incident is shown in FIG. 10. The
refractive indexes line function of positive incident is in FIG.
5(a), and then the refractive indexes line function of negative
incident is in FIG. 5(b). The FIG. 5(a) is the function of line
increasing, and the FIG. 5(b) is the function of line decreasing.
From Eq. (37), we can calculate the electric field distribution of
light in the GFPCs. For the positive incident, the electric field
distribution is shown in FIG. 11(a), and FIG. 11(b) is the
electric field distribution for the negative incident. From FIG.
11(a) and (b), we can obtain the result: (1) when the refractive
indexes function increase at the incident direction of light, the
light intensity should be boosted up or magnified, which can be
made light magnifier (magnifying multiple $10^{6}$). (2) when the
refractive indexes function decrease at the incident direction of
light, the light intensity should be weaken, which can be made
light attenuator (magnifying multiple less than $1$). Actually,
the GFPCs structure in FIG. 10 is the optical diode, since the
light intensity be magnified as the light positive incident, and
the light intensity should be decreased as the light negative
incident.

\vskip 5pt

{\bf 6. Conclusion}

\vskip 8pt

In summary, We have theoretically investigated a new general
function photonic crystals (GFPCs), which refractive index is a
function of space position. Based on Fermat principle, we achieve
the motion equations of light in one-dimensional general function
photonic crystals, and calculate its transfer matrix. We choose
the line refractive index function for two mediums $A$ and $B$,
and obtain some results: (1) when the line function of refractive
indexes is up, the transmissivity can be far larger than $1$. (2)
when the line function of refractive indexes is down, the
transmissivity can be far smaller than $1$. (3) when the
refractive indexes function increase at the direction of incident
light, the light intensity should be magnified, which can be made
light magnifier. (4) when the refractive indexes function decrease
at the direction of incident light, the light intensity should be
weaken, which can be made light attenuator. (5) The GFPCs can be
made light diode. The new general function photonic crystals can
be applied to design more optical instruments.
\\

\end{document}